\documentclass[aps,prl,letterpaper,10pt,twocolumn,showpacs,superscriptaddress]{revtex4-1}

\usepackage{lipsum}

\usepackage{amsfonts,amssymb,amsmath,amsthm,graphicx}
\usepackage{gensymb}
\usepackage{url}
\usepackage{dsfont}
\usepackage{bbm}
\usepackage[usenames,dvipsnames]{xcolor}
\usepackage{nicefrac}
\usepackage{natbib}
\usepackage{setspace}
\usepackage{subfigure}
\usepackage{comment}
\usepackage{enumitem}
\setlist[itemize]{noitemsep, nolistsep}
\usepackage{natbib}

\usepackage{physics}
\usepackage{amsmath}
\usepackage{graphicx}
\usepackage[colorinlistoftodos]{todonotes}
\usepackage[colorlinks=true, allcolors=blue]{hyperref}
\usepackage{color, colortbl}
\definecolor{Gray}{gray}{0.9}

\usepackage{times,todonotes}
\usepackage[normalem]{ulem}

\def\virgolette #1{``#1"}

\newcommand{\beq}{\begin{equation}}
\newcommand{\eeq}{\end{equation}}

\begin{document}
\raggedbottom

\title{Full daylight quantum-key-distribution at 1550~nm enabled by integrated silicon photonics}

\author{M.~Avesani}
\author{L.~Calderaro}
\author{M.~Schiavon}
\author{A.~Stanco}
\affiliation{Dipartimento di Ingegneria dell'Informazione, Universit\`a di Padova, via Gradenigo 6B, 35131 Padova, Italy}
\author{C.~Agnesi}
\author{A.~Santamato}
\author{M. Zahidy}
\author{A.~Scriminich}
\author{G.~Foletto}
\affiliation{Dipartimento di Ingegneria dell'Informazione, Universit\`a di Padova, via Gradenigo 6B, 35131 Padova, Italy}
\author{G.~Contestabile}
\affiliation{Istituto TeCIP - Scuola Superiore Sant'Anna, Pisa Italy}
\author{M.~Chiesa}
\author{D. ~Rotta}
\affiliation{InPhoTec, Integrated Photonic Technologies Foundation, Pisa Italy}
\author{M.~Artiglia}
\author{A.~Montanaro}
\author{M.~Romagnoli}
\author{V.~Sorianello }
\affiliation{PNTLab - Consorzio Nazionale Interuniversitario per le Telecomunicazioni, Pisa Italy}
\author{F.~Vedovato}
\affiliation{Dipartimento di Ingegneria dell'Informazione, Universit\`a di Padova, via Gradenigo 6B, 35131 Padova, Italy}
\author{G.~Vallone}
\affiliation{Dipartimento di Ingegneria dell'Informazione, Universit\`a di Padova, via Gradenigo 6B, 35131 Padova, Italy}
\affiliation{Dipartimento di Fisica e Astronomia, Universit\`a di Padova, via Marzolo 8, 35131 Padova, Italy}
\author{P.~Villoresi}
\email{paolo.villoresi@dei.unipd.it}
\affiliation{Dipartimento di Ingegneria dell'Informazione, Universit\`a di Padova, via Gradenigo 6B, 35131 Padova, Italy}

\date{\today}

\begin{abstract}
The future envisaged global-scale quantum communication network will comprise various nodes interconnected via optical fibers or free-space channels, depending on the link distance.
The free-space segment of such a network should guarantee certain key requirements, such as daytime operation and the compatibility with the complementary telecom-based fiber infrastructure. In addition, space-to-ground links will require the capability of designing light and compact quantum devices to be placed in orbit. For these reasons, investigating available solutions matching all the above requirements is still necessary. Here we present a full prototype for daylight quantum key distribution at 1550~nm  exploiting an integrated silicon-photonics chip as state encoder. We tested our prototype in the urban area of Padua (Italy) over a 145m-long free-space link, obtaining a quantum bit error rate around 0.5\% and an averaged secret key rate of 30~kbps during a whole sunny day (from 11:00 to 20:00). The developed chip represents a cost-effective solution for portable free-space transmitters and a promising resource to design quantum optical payloads for future satellite missions.
\end{abstract}

\maketitle

\section{Introduction}

Quantum Key Distribution (QKD)~\cite{GisinQKD,ScaraniSecurityQKD,DiamantiQKD} is the most advanced application of quantum information science, continuously improving in terms of new  protocols~\cite{Rusca2018,Lucamarini2018} and experimental realizations~\cite{BoaronRecord,MinderPittaluga2019,Pognac19,2019arXiv190206268L}. The potential of QKD is to allow secure communication between any two points on Earth. Depending on the link distance, the quantum channel is established through fiber-based or free-space quantum communication (QC).

Despite the recent demonstrations also realized in satellite-to-ground links~\cite{Liao2017,Bedington2017,Agnesi2018,Khan2018}, free-space QKD-technology is currently limited and cannot compete with its fiber-based counterpart~\cite{BoaronRecord,Yoshino2013,Islam2017,Yuan2018}. Hence, in the vision of a continental-scale quantum network (or quantum internet)~\cite{Peev2009,Sasaki2011,Kimble08,Wehnereaam9288}) where both types of link are required to jointly operate, certain key requirements for free-space QC can be formulated, as i) full-day functionality, ii) compatibility with standard fiber-based technology at telecom wavelength, and iii) the achievement of stable coupling of the free-space signal into a single-mode fiber (SMF).

Regarding i), the background noise due to sunlight poses a serious limitation on the achievable performance of day-time free-space QC, limiting most of the demonstrations obtained so far to night-time. For this reason, various studies have focused on the feasibility of {\it daylight} QKD~\cite{Buttler2000,Hughes2002,Peloso2009,Ko2018,Liao2017_daylight,Gong2018}. Most of them exploited light in the 700-900 nm band which allows for for a good atmospheric transmission, and to exploit commercial low noise silicon-based single-photon avalanche diodes (SPADs). To reduce the background noise due to Sun and to maintain, at the same time, a good efficiency in the atmospheric transmission, the choice to use light signals in the telecom C-band (around 1550~nm) has only very recently started to be investigated~\cite{Liao2017_daylight,Gong2018}.

Moving the operating wavelength to the telecom band comes with (at least) two advantages. Firstly, it is the standard choice in fiber-based optical (classical and) QC, hence fulfilling the requirement ii). Secondly, it is compatible with integrated silicon photonics~\cite{Ma:16,Sibson2017,Bunandar2018,Agnesi2019}, which represents a promising choice for designing light, compact, scalable and low power-consuming  devices suitable for portable QKD transmitter and to design satellite optical payloads~\cite{SPIE.NASA,siph.space}. 

Furthermore, to match the requirement iii) it is necessary to actively compensate for the optical aberrations (at least the beam wander and angle-of-arrival fluctuation) introduced by atmospheric turbulence, which is experimentally challenging. However, a stable coupling of the light signal into a SMF has the advantage of allowing the use of commercially available superconductive nanowire single-photon detectors (SNSPDs), which represents the standard for fiber-based state-of-the-art QKD demonstrations~\cite{BoaronRecord,2019arXiv190206268L,Islam2017}.

\begin{figure*}[t]
    \centering
    \includegraphics[width=0.99\textwidth]{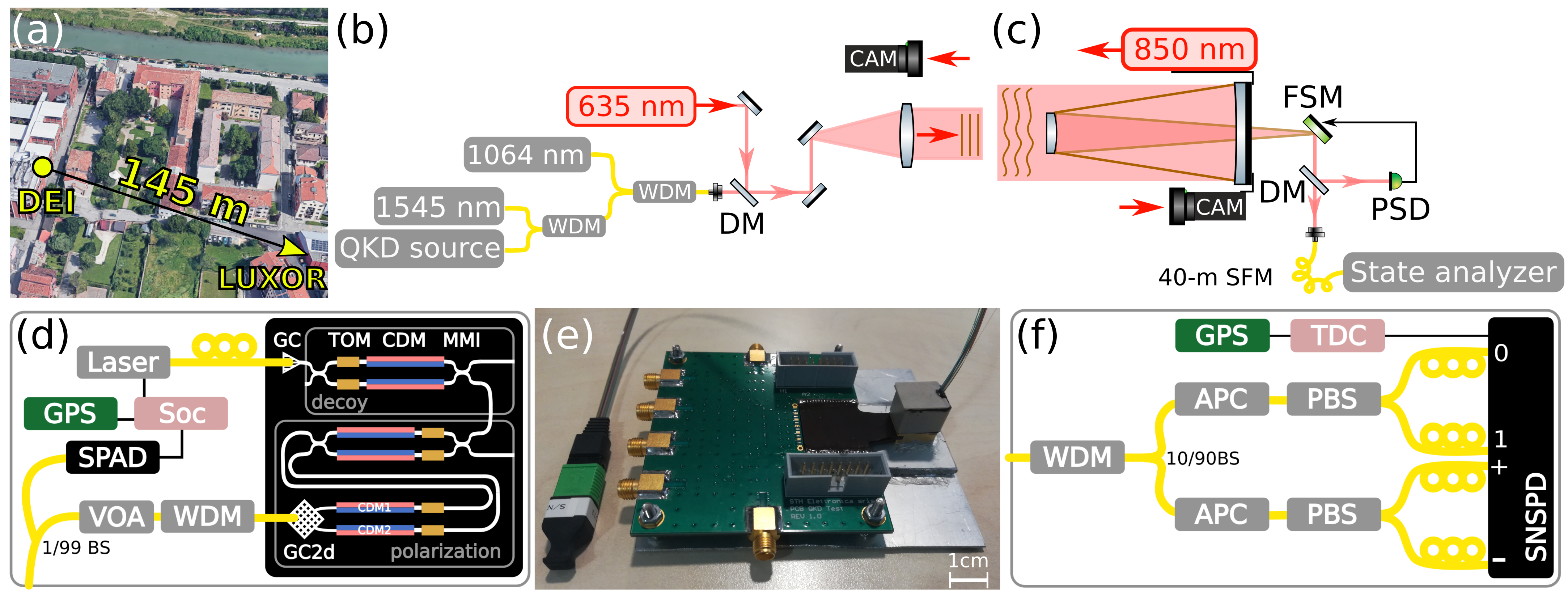} 
    \caption{{\it Location of the field trial and QCoSOne setup.}  (a) Location of the field test in Padua. Map Data from Google --- \copyright2019 Google. (b) Alice's terminal. DM: dichroic mirror; CAM: camera. (c) Bob's terminal. (d) Scheme of Alice's QKD source with chip schematic. (e) Picture of the packaged chip soldered to the external control board with the input/output fiber array. (f) Scheme of Bob' state analyzer.}
    \label{fig:link}
\end{figure*}

Here we address the requirements outlined above presenting a QC-system named {\it QCoSOne} (acronym for \virgolette{Quantum Communication for Space-One}), which realizes free-space daylight QKD at 1550~nm.
We exploited integrated silicon-photonics technology to realize a portable state encoder with decoy- and polarization-modulation on a single chip, to implement the 3-state 1-decoy QKD protocol introduced in Ref.~\cite{Rusca2018}. Moreover, the integrated QKD encoder has been put in a  
rugged package, designed and realized to the purpose in house, making  
it ready for use in the field. We exploited commercially available wavelength filters (with~$\lesssim$1~nm of bandwidth) and SNSPDs at the detection state. We reached a stable SMF coupling of the qubit stream over a 145m-long free-space link [Fig.~\ref{fig:link}(a)] by means of an active correction of the first order aberrations that allowed us to successfully perform QKD in daylight continuously from 11:00 to 20:00 (Central European  Summer Time, CEST). We measured a quantum bit error rate (QBER) around 0.5\% in most of the runs, obtaining a final secret-key-rate (SKR) up to 65 kbps after finite-key analysis.  To the best of our knowledge, this is the highest secret key rate obtained in a free-space QKD demonstration at 1550~nm, and the first daylight experiment where a chip-based QKD state encoder is coupled to a free-space channel in an actual field trial. Our result represents one step  towards the demonstration of a seamless joint satellite-fiber quantum network at the telecom band.  

\section{Results}

\subsection{Description of QCoSOne prototype}

 A comprehensive sketch of the experimental setup is presented in Fig.~\ref{fig:link}. The QKD source and detection units, at Alice's and Bob's side respectively, are linked by a free-space channel established between the Department of Information Engineering (DEI) and the Luxor Laboratory (LUXOR) [Fig.~\ref{fig:link}(a)]. Alice's transmitter telescope is an achromatic refractor with an aperture diameter of 12~cm, mounted on a motorized tripod and placed at DEI building [Fig.~\ref{fig:link}(b)]. Bob's receiving telescope is a Cassegrain reflector with an equivalent focal length of 5000~mm comprising a primary mirror with a diameter of 31.5~cm, mounted on a fixed support located at LUXOR [Fig.~\ref{fig:link}(c)].

We focused and coupled the  free-space signal beam down to the 14~$\mu$rad of field-of-view (FOV) of the receiver. In order to do this, we developed a closed-loop feedback control at Bob's side, based on the use of a fast-steering mirror (FSM) [SmarAct] to correct for lower-order aberrations (i.e., tip-tilt) induced by turbulence. The feedback signal is provided by the centroid of an auxiliary beacon laser at 1064~nm acquired by a position sensitive detector (PSD). We reached an average coupling efficiency into  the SMF of around 4\% (14 dB), which is in line with analogous systems based only on tip-tilt correction~\cite{Liao2017_daylight}. More details on the pointing, acquisition and tracking (PAT) system are presented in the Methods section.

{\it QKD source and chip encoder.---}Fig.~\ref{fig:link}(d) shows the sketch of the QKD source. A gain-switched distributed feedback (DFB) laser source outputs a 50~MHz stream of phase-randomized pulses (with 500~ps of full-width-at-half-maximum, FWHM) at 1550~nm of wavelength.
Such pulses are coupled in and out of a state encoder realized in a silicon-based photonic integrated circuit (PIC) via a standard 8-channel SMF-array glued to the PIC's grating couplers (GCs). The PIC was designed in-house and realized exploiting the Europractice IC Service~\cite{EUROPRACTICE} offered by the IMEC foundry. 

The PIC comprises several interferometric structures exploiting standard building blocks provided by the foundry, e.g. multi-mode interference (MMI) devices acting as 50/50 beam splitters, slow thermo-optics  modulators (TOMs, $\sim$kHz of bandwidth, DC modulation) and fast carrier-depletion modulators (CDMs, $\sim$GHz of bandwidth, RF modulation), which are realized in a reverse biased $p$-$n$ junction on the silicon waveguide. More details on the working principle of the components of the PIC and on the fabrication process can be found in Refs.~\cite{Velha:16,Sibson2017,IMEC_absil}. 

The size of the PIC is about 5 mm $\times$ 5 mm, while the complete package is compact within a total volume of 1.2 cm $\times$ 1.5 cm $\times$ 1.2 cm.  A picture of the packaged PIC soldered to a standard 7 cm $\times$ 8 cm control board is presented in Fig.~\ref{fig:link}(e). The package has been designed, developed and assembled to the purpose in-house (featuring 20 DC and 6 RF ports), so as to make  it rugged, portable and easily usable in  
field experiments. The capability of compacting and integrating the devices needed to prepare the quantum states down to such a small volume represents an attractive choice for designing portable transmitters and payloads for satellite QC~\cite{Oi2017}. 

The first interferometric structure realizing a Mach-Zehnder interferometer (MZI) is used to obtain the amplitude modulation of the pulses according to the chosen QKD protocol (see Ref.~\cite{Rusca2018} and Methods for more details), which requires preparing pulses with two different intensity levels, or mean photon number per pulse $\mu_1$ and $\mu_2$, with $\mu_1 > \mu_2$. The ratio $\mu_1/\mu_2$ is set by both  the DC-bias of the TOMs as well as by the RF signal amplitude sent to the CDMs of the MZI. In particular, if no RF signal is applied, the amplitude is set to $\mu_1$, while, by applying an RF signal, the $\mu_2$ intensity is sent. 

The second structure allows to realize the polarization modulation, by exploiting an inner MZI, followed by two external CDMs [CDM1 and CDM2 in Fig.~\ref{fig:link}(d)] ending in a 2-dimensional grating coupler (GC2d). The GC2d converts the path-encoded information used within the PIC into the polarization-encoded information at the output-SMF. 
Referring to the Bloch sphere, the colatitude $\theta$ of the produced polarization state $\ket{\psi} = \cos{(\theta/2)}\ket{H} + e^{i \varphi} \sin{(\theta/2)} \ket{V}$ is controlled by acting on the internal MZI, whereas the longitude $\varphi$ is set by acting on the external phase modulators. Therefore, by voltage biasing the TOMs of the inner MZI the balanced superposition of horizontal and vertical polarization $\ket{+} = (\ket{H} + \ket{V} )/\sqrt{2}$ is created. If no RF signal is applied, the output state remains $\ket{+}$, whereas, by applying an RF signal on the external CDMs, a $\pi/2$ phase shift can imposed to either arm, respectively creating the states $\ket{L} = (\ket{H} - i \ket{V})/\sqrt{2}$ or $\ket{R} = (\ket{H}  + i \ket{V})/\sqrt{2}$. In this way, we obtained the three states required by the protocol for the key-generation basis $\mathcal{Z} = \{\ket{0}, \ket{1}\}$, where $\ket{0}:=\ket{L}$, $\ket{1}:= \ket{R}$, and the control basis $\mathcal{X} = \{\ket{+},\ket{-}\}$, with $\ket{\pm} :=(\ket{H}\pm \ket{V})/\sqrt{2}$.
Each produced state is characterized by an extinction ratio (ER), i.e. the ratio between the optical power in two orthogonal polarizations, up to 30~dB. 

The pulses exiting from the PIC are spectrally filtered by a commercial dense wavelength-division-multiplexing (WDM) filter with 100~GHz of channel spacing, corresponding to a bandwidth of 0.8~nm centered around 1550.94~nm. Then, they pass through a variable optical attenuator (VOA) to set $\mu_1$ to the desired level, and a 99/1 fiber beam splitter (BS). The 1\% output is directed to an InGaAs/InP gated SPAD [Micro Photon Devices~\cite{Tosi2012}] to monitor in real time the intensity level of the pulses, while the 99\% output is directed to the transmitting telescope together with the additional beacon lasers at 1064~nm and 1545~nm used by the PAT system.

We made Alice generate the polarization pulses with basis-probability $p_A^\mathcal{Z} = 0.9$ and $p_A^\mathcal{X} = 0.1$ and intensities $\mu_1^{\mathcal{Z}}   =  0.56$, $\mu_2^{\mathcal{Z}}  = 0.27$, $\mu_1^{\mathcal{X}} = 0.69$, and $\mu_2^{\mathcal{X}} = 0.33$ at the aperture of the transmitting telescope, with decoy-probability $p_{\mu_1} = 0.7$ and $p_{\mu_2} = 0.3$. These working parameters are close to optimal for a total attenuation ranging from 20 to 30~dB, a QBER of the order of 1\% and a number of sifted bits $n_{\mathcal{Z}} \gtrsim 10^8$, as we expected in our experiment according to our simulations and Ref.~\cite{Rusca2018}.
It is worth noting that the possibility of using different intensity levels for the two bases without losing security (as discussed in Ref.~\cite{Yu2016}) is particularly interesting when dealing with non ideal CDMs, since they typically incur phase-dependent losses translating into polarization-dependent amplitude levels of the QKD pulses.

 The random bits used for running the protocol are obtained from the source-device-independent quantum random number generator based on the heterodyne measurement of the electromagnetic field described in~\cite{Avesani2018}. An evaluation board [ZedBoard by Avnet] with a System-on-a-chip (Soc) is used to control the source. The board clock is locked to a 10~MHz clock from Alice's GPS module. The field-programmable-gate-array (FPGA) side of the Soc is designed in order to produce the electrical pulses for triggering the laser, the RF signals driving the CDMs and the gating signal for the SPAD.

{\it State analyzer.---}Bob's SMF-based state analyzer in Fig.~\ref{fig:link}(f) is connected to the receiving telescope by a 40m-long SMF. The state analyzer comprises first a dense WDM (matched with the one of the source) to select the photons coming from Alice, and then a 90/10 fiber BS setting the detection probabilities of the two measurement bases to $p^\mathcal{Z}_B = 0.9$ and $p^\mathcal{X}_B = 0.1$. Each output arm of the BS is  connected to an automatic polarization controller (APC) and a polarizing beam splitter (PBS). The four outputs are sent to four SNSPDs
[ID281 by ID-Quantique]
cooled to 0.8~K. A manual polarization controller at the input of every detector is used to optimize the detection efficiency. This is around 85\% for the detectors in the $\mathcal{Z}$ basis, whereas it is 90\% and 30\% for the $\ket{+}$ and $\ket{-}$ detectors, respectively. As discussed in Refs.~\cite{Fung2009,Bochkov2019}, we randomly discarded some detections in post-processing in order to balance the different efficiencies. 
All the detectors are affected by about 200~Hz of intrinsic dark count rate.

The SNSPD detections and the pulse-per-second (PPS) signal produced by the GPS module located at Bob' side are recorded by a time-to-digital converter (TDC) [qutools] with 81~ps of temporal resolution. In order to time-correlate the two terminals, we developed a self-synchronization system which is based only on the use of the two GPS modules.

\subsection{The field trial}

Exploiting QCoSOne, we performed multiple QKD runs during the month of April 2019, on several days of clear sky condition. After aligning the two telescopes, and reaching a good SMF coupling efficiency, we aligned the two measurement bases at Bob's side to $\mathcal{Z}$ and $\mathcal{X}$ by making Alice send a fixed polarization pattern. Exploiting the APCs in the state analyzer, an ER above 20 dB was obtained for all the polarization states.

\begin{figure}[t]
\centering
\includegraphics[width=0.45\textwidth]{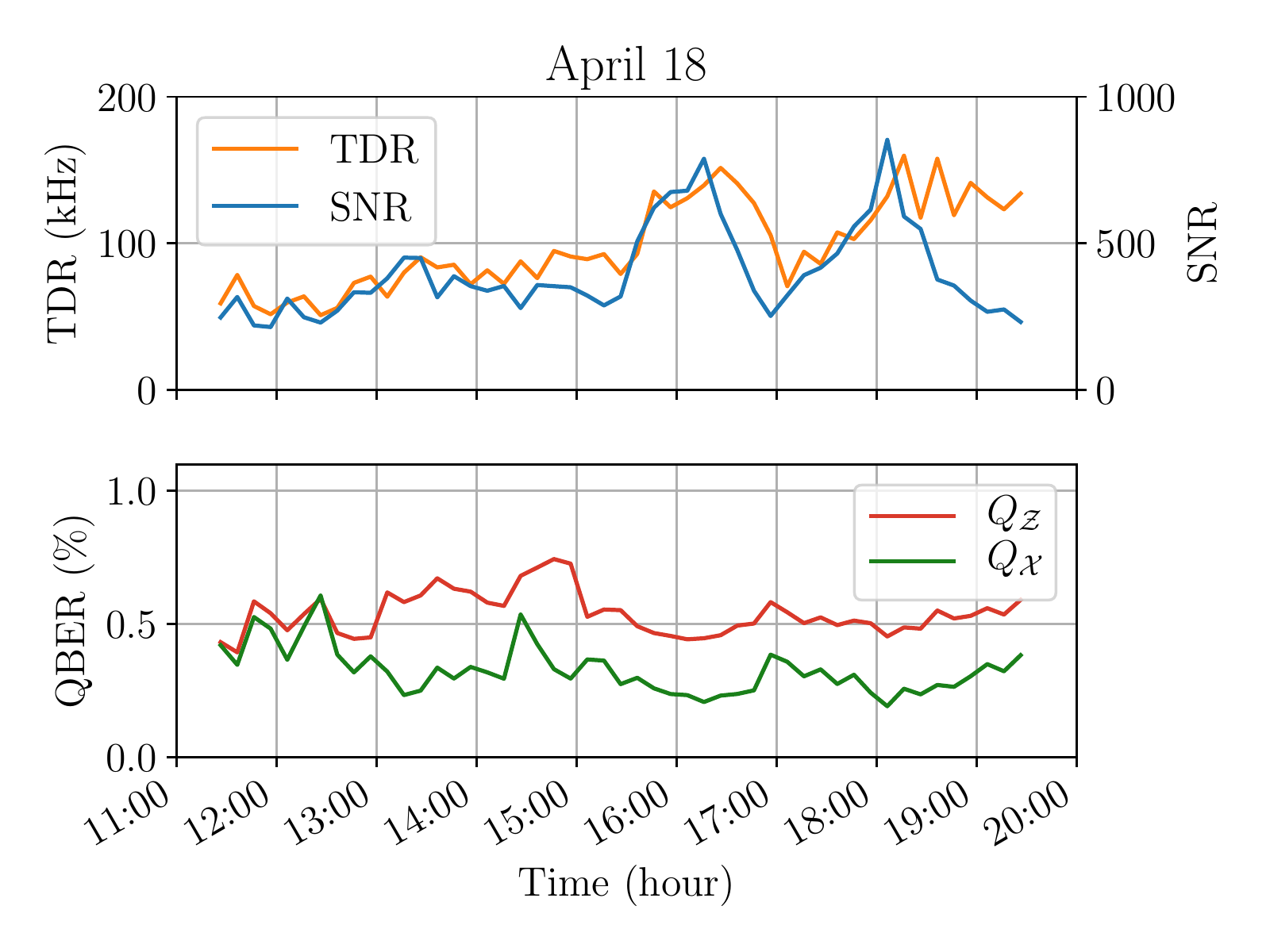}
\caption{{\it TDR, SNR and QBER obtained on April 18th, 2019.}} \label{fig:only_18}
\end{figure}

On April 18th we managed to perform the QKD experiment continuously for eight hours of daylight, as shown in Fig.~\ref{fig:only_18}. The total detection rate  (TDR, orange line) within a 1ns-wide detection window around the expected arrival time of the pulses (when renormalized taking into account the different quantum efficiencies of the SNSPDs) ranges from 60 to 130~kHz, being around 100~kHz on average.  As expected, in our experiment the SMF coupling efficiency and hence the TDR increased approaching the late afternoon, thanks to the reduced turbulence due to the weaker temperature gradient. In daylight, the background rate within the detection window due to environmental light varies, ranging from 200 to 400~Hz and being about 240~Hz on average. Hence, the  signal-to-noise ratio (SNR, blue line) is about 400, while the total losses are around 24~dB on average (5~dB of fixed attenuation due to the optics of the receiver, 5~dB of fixed attenuation in the state analyzer and  14~dB due to the mean SMF coupling efficiency). The drop of the SNR after 18:30 is due to the fact that the receiving telescope was facing toward the sunset, hence increasing the background rate.

 We notice that, by narrowing the detection window, the SNR increases (at the expense of a lower sifted rate). In our case, by reducing the detection window to 500~ps, the detection rate decreases by 25\%, while the noise is reduced by 50\%. For low SNR values, the above strategy may result in a higher key rate~\cite{Vallone2015pra}.
With the reduced detection window, the simulation of the post-processing procedure (see Methods) provides that our setup would be able to produce a secret key even with 14 dB of additional losses
(if only beam-diffraction is considered such losses would correspond to a link distance of about 50 km).

The measured QBER is less than 0.75\% for all of the eight hours without the use of any active polarization stabilization system,  reaching a value as low as $Q_{\mathcal{Z}} \approx 0.45\%$ in the $\mathcal{Z}$ basis and $Q_{\mathcal{X}} \approx 0.25\%$ in the $\mathcal{X}$ basis. This is the best result to date for a free-space QKD system operating in daylight~\cite{Buttler2000,Hughes2002,Peloso2009,Ko2018,Liao2017_daylight,Gong2018}, with performances comparable to fiber-based systems~\cite{Yoshino2013,Yuan2018}. This result demonstrates that the developed chip encoder is characterized by an excellent polarization stability over time.
This feature makes silicon-photonics PICs very attractive in the context of polarization-based satellite QC.

\begin{figure*}[t]
    \centering
    \includegraphics[width=\textwidth]{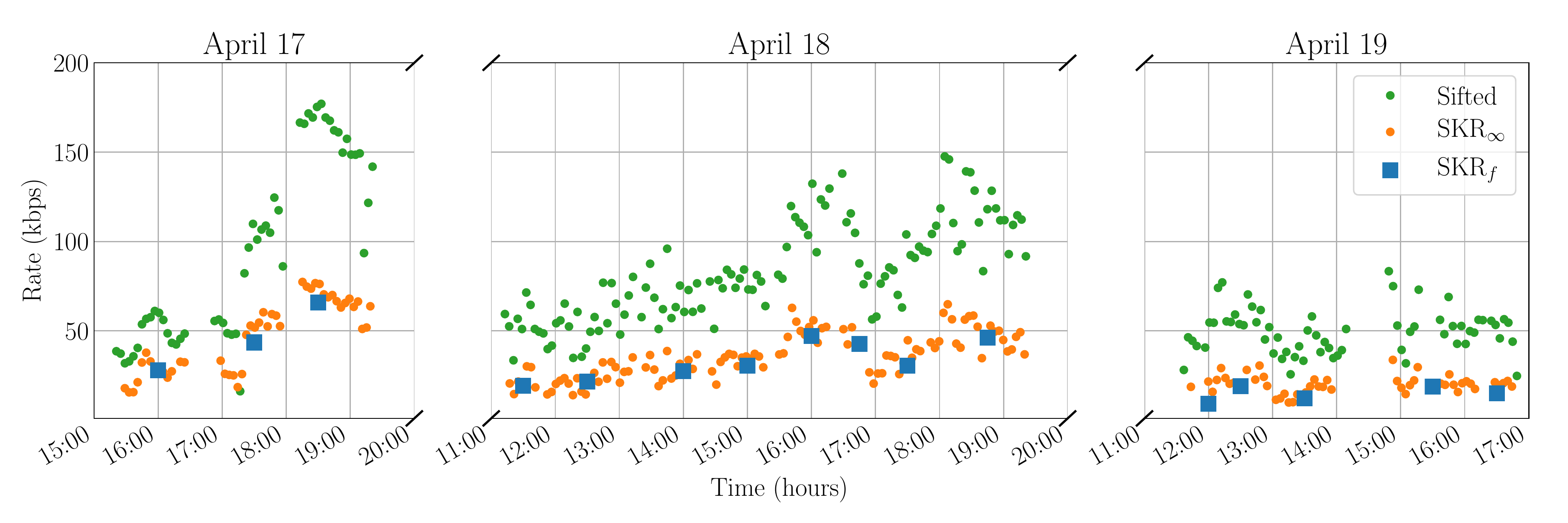} 
    \caption{ {\it Results obtained in daylight during three consecutive days of QKD runs.} The maximum Sun elevation was about 55$\degree$ at 13:00; sunset was around 20:00.}
    \label{fig:alldays}
\end{figure*}

In Fig.~\ref{fig:alldays} we report the results of the different QKD runs performed over three consecutive days. The weather conditions were  good on all of the three days, with a clear and sunny sky. The Sun reached its maximum elevation (55$\degree$) around 13:00 and the sunset was around 20:00. Each QKD run lasted for the time needed to guarantee that the requirement $n_{\mathcal{Z}} \gtrsim 10^8$ was fulfilled. As we showed in Fig.~\ref{fig:only_18}, the TDR increased during the day, thus making the effective duration of the QKD runs vary, typically from 15 to 55 minutes.   

Each graph in Fig.~\ref{fig:alldays} shows the rate of the sifted bits $n_{\mathcal{Z}}$ (green dots), the asymptotic (infinite-size) SKR ($\mathrm{SKR}_\infty$, orange dots) and the finite-size SKR ($\mathrm{SKR}_f$, blue squares, see Methods for more details) as a function of the hour of the day. Each dot is obtained by an average over four minutes of data acquisition by merging all the runs, while each $\mathrm{SKR}_f$ point is obtained with a single QKD run. 
The obtained results are comparable over the three days. The sifted bit rate ranges from 50 to 150~kbps, depending essentially on the TDR, hence showing an improvement while approaching the late afternoon. The same trend characterizes also the $\mathrm{SKR}_\infty$, which ranges from 20 to 70~kbps. We manged to obtain a $\mathrm{SKR}_f$ of several tens of kbps for all days, reaching a maximum of 65.8~kbps in the last acquisition of April 17th. Remarkably, each QKD run performed on April 18th lasted for about 50 minutes, allowing to obtain a mean $\mathrm{SKR}_f$ about 33 kbps, hence outperforming the results obtained with comparable free-space QKD system at 1550~nm by two orders of magnitude~\cite{Liao2017_daylight,Gong2018}.     

It is worth noticing that a complete QKD experiment in free-space  has never before been performed with the Sun at its maximum elevation. Indeed, Gong {\it et al.} in~\cite{Gong2018} tried to perform QKD for the whole daytime in a 8km-long link in Shanghai, but the impracticable turbulence conditions and the sunlight background did not allow them to extract a key at around noon. We demonstrated that performing daylight QKD in the middle of the day (around 13:00 in our case) is possible, obtaining a SKR of tens of kbps even in such a condition in two different days. 

\section{Discussion}

In conclusion, we demonstrated the successful realization of a chip-based prototype for free-space QC in daylight, operating at a wavelength in the telecom C-band. We performed several QKD runs obtaining an extremely low QBER ($\sim$0.5\%) and a SKR of several tens of kbps, also with the Sun at its maximum elevation. We overcame the strong background noise coming from the Sun light by  exploiting temporal (i.e., synchronization), spatial (i.e., single mode fibers) and wavelength (i.e, dense WDM) filters. To our knowledge, this is the first time that intensity and polarization modulations are realized in a single chip used as qubit encoder for decoy-state QKD, as well the first time that such integrated technology is used in a real free-space QKD-trial in an urban area, thanks to the dedicated packaging designed  
and realized to the purpose. Our solution is very attractive for the design and development of optical payloads to be placed in portable terminals or satellites dedicated to QC, given the low resources needed in terms of power, weight and space. 

Further improvements to our prototype can be achieved by increasing the system clock rate, for example up to 1~GHz (as in Refs.~\cite{BoaronRecord,Yuan2018,Sibson2017,Bunandar2018}), and exploiting adaptive optics to increase the SMF coupling efficiency~\cite{Chen2015} and thus the tolerable losses and achievable link distance. However, the obtained results show that daylight QKD technology is mature enough to foresee the real application of a global scale QC-network in the next future~\cite{Kimble08,Wehnereaam9288}. It will likely comprise free-space, satellite and fiber-based channels exploiting  quantum technologies
to accomplish tasks such as QKD, realizable also in the device- or measurement-device independent framework~\cite{Acin2007,Liu2019}, entanglement distribution~\cite{Yin2017}, quantum teleportation~\cite{Ren2017} and quantum time distribution~\cite{Komar2014}, as envisaged by the Italian Quantum Backbone~\cite{Calonico2016}, a fiber-based infrastructure connecting the National Institute of Metrological Research in Turin with the ASI Space Center in Matera.

\section{Methods} 

\subsection{Description of the QKD protocol}
In our  prototype  we chose to realize the 3-state 1-decoy version of the  efficient BB84 protocol proposed by Rusca {\it et al.}~\cite{Rusca2018}, due to the reduced complexity of the implementation when exploiting polarization encoding. The protocol works as follows.

Alice randomly encodes a weak coherent pulse either in the  $\mathcal{Z} =\{\ket{0}, \ket{1}\}$ basis, with probability $p^{\mathcal{Z}}_A$, or in the $\mathcal{X} = \{\ket{+}, \ket{-}\}$ basis, with probability $p^\mathcal{X}_A = 1 - p^\mathcal{Z}_A$. 
The basis $\mathcal X$ is Mutually
Unbiased with respect to $\mathcal Z$, namely
$|\braket{0}{\pm}|^2=|\braket{1}{\pm}|^2=1/2$. In our implementation, we have chosen $\ket{0}:=\ket L = (\ket{H}  -  i \ket{V})/\sqrt{2}$, $\ket{1}:=\ket R = (\ket{H}  +  i \ket{V})/\sqrt{2}$ and $\ket{\pm} := (\ket{H} \pm \ket{V})/\sqrt{2}$.
Alice needs to generate only three polarization states, $\ket{0}$ and $\ket{1}$ with uniform probability for the $\mathcal{Z}$ basis, and $\ket{+}$ for the $\mathcal{X}$ one.
The intensity level of the pulse is randomly chosen between two values, $\mu_1$ and $\mu_2$, with probabilities $p_{\mu_1}$ and $p_{\mu_2} = 1- p_{\mu_1}$, respectively. The two values can differ between pulses prepared in $\mathcal{X}$ and $\mathcal{Z}$ ($\mu_1^{\mathcal{X}} \neq \mu_1^{\mathcal{Z}}$, $\mu_2^{\mathcal{X}} \neq \mu_2^{\mathcal{Z}}$), because we carry out the yield analysis separately in the two bases \cite{Yu2016}. This procedure allows to detect a possible photon-number-splitting attack~\cite{PNS}.  

Bob, at his site, measures the incoming photons in the two bases $\mathcal{Z}$ and $\mathcal{X}$, with probability $p^\mathcal{Z}_B$ and $p^\mathcal{X}_B = 1 - p^\mathcal{Z}_B$, respectively. After the photons exchange, Alice and Bob announce, for each detected event, their basis choices. Then, $n_\mathcal{Z}$ raw sifted bits are obtained by  comparing the detections in the $\mathcal{Z}$ basis, while the ones from the $\mathcal{X}$ basis are used to estimate the information leakage toward a potential eavesdropper. 

After generating a raw key, Alice and Bob proceed with the error correction (EC) and the finite-key privacy amplification (PA) steps, ultimately obtaining, for each PA block, a secure secret key of $l$ bits, which is bounded by~\cite{Rusca2018}:
\begin{align}
    l &\leq s_{\mathcal{Z},0} + s_{\mathcal{Z},1}(1 - h(\phi_\mathcal{Z})) - \lambda_{\rm EC} \nonumber\\   &\quad\quad  - 6 \log_2(19/\epsilon_{\rm sec}) - \log_2(2/\epsilon_{\rm cor}) \ ,
\end{align}
where $s_{\mathcal{Z},0}$ and $s_{\mathcal{Z},1}$ are the lower bounds on the number of vacuum and single-photon detections in the $\mathcal{Z}$ basis, $\phi_\mathcal{Z}$ is the upper bound on the phase error rate corresponding to single photon pulses, $h(\cdot)$ is the binary entropy, $\lambda_{\rm EC} = f_{\rm EC} n_\mathcal{Z} h (Q_\mathcal{Z})$ is the total number of bits revealed during the EC step --- which depends on the reconciliation efficiency of the EC algorithm (Cascade, in our case $f_{\rm EC} \approx  1.06$), the number of raw key bits $n_\mathcal{Z}$, and on the QBER $Q_\mathcal{Z}$ --- and $\epsilon_{\rm sec} =  10^{-10}$, $\epsilon_{\rm cor} = 10^{-12}$ are the secrecy and the correctness parameters, respectively~\cite{Rusca2018}. 
The results for the different QKD runs presented in this work were obtained by adapting the AIT QKD R10 software suite by the AIT Austrian Institute of Technology GmbH~\cite{AIT} to our needs. 

\subsection{Additional details on the PAT system}

The complete setup comprising the PAT system is presented in Fig.~\ref{fig:link}(b) and Fig.~\ref{fig:link}(c), and it is based on a \textit{coarse-alignment} system and on a \textit{fine-alignment} one, in charge of the SMF coupling. The former is based on the use of two counter-propagating beacon lasers (at 635~nm and 850~nm of wavelength) which are acquired by a CMOS camera at both terminals to guarantee a rough alignment between the two telescopes. The fine alignment is achieved by controlling in closed-loop the FSM with the feedback signal provided by the PSD acquiring the 1064-nm beacon laser sent by Alice, as described in the main text. The coupling efficiency is estimated by monitoring the power into the SMF collected from an auxiliary beacon at 1545~nm of wavelength. 

The performance of the PAT system has been evaluated by looking at the mean coupling efficiency with and without the tip-tilt correction in the 145-m link. The mean coupling efficiency is increased from an average of $\sim$1\% without control to $\sim$4\% with the tip-tilt correction, with peak values up to 10\%. This value is lower than the 20\% measured in the laboratory, due to the turbulence in the optical link. 
Our results are comparable with similar systems based only on tip-tilt correction, as in~\cite{Liao2017_daylight}.

\section{Acknowledgements}
\begin{acknowledgements}
We acknowledge the Italian Space Agency for support, and specifically Dr. Alberto Tuozzi and Dr. Claudia Facchinetti of the Telecommunication and Navigation division. This work was funded by the project QCommSpaceOne of the Italian Space Agency (ASI, Accordo n. 2017-4-H.0): \virgolette{Space Quantum Communications between Space and Earth: conception of a flight terminal and development of a prototype}. We thank Iyad Suleiman for his contribution in the early development of the fiber-injection system. We thank Lorenzo Franceschin and
Achille Forzan of the Department of Information Engineering of the University of Padua for logistic support to realize the field trial.
We thank Christoph Pacher, Oliver Maurhart and the AIT Austrian Institute of Technology GmbH for providing the foundation for the post-processing software.
\end{acknowledgements}

\bibliography{qcosone}
\end{document}